# A temporal access code to consciousness?

Birgitta Dresp-Langley
CNRS UMR 7357
Strasbourg, FRANCE

Short title: A brain code to consciousness

Keywords: Conscious States – Biophysical Time – Resonant Brain Learning – Temporal Activity Patterns – Time-Bin Coding

## Abstract

While questions of a functional localization of consciousness in the brain have been the subject of myriad studies, the idea of a temporal access code as a specific brain mechanism for consciousness has remained a neglected possibility. Dresp-Langley and Durup (2009; 2012) proposed a theoretical approach in terms of a temporal access mechanism for consciousness based on its two universally recognized properties. Consciousness is limited in processing capacity and described by a unique processing stream across a single dimension: time. The time ordering function of conscious states is highlighted and neurobiological theories of the temporal brain activities likely to underlie such function are reviewed. Arguments for a purely temporal access code for conscious states are discussed, including Ramachandran's 'Remapping Hypothesis', research on the 'Coherence Index' and coincidence detectors, and theoretical models of adaptive resonant matching of bottom-up and top-down representations. We conclude that a purely temporal resonance mechanism provides the most parsimonious neurobiological explanation of consciousness and propose a '*time-bin resonance model'*, where temporal messages for conscious state access are generated on the basis of signal reverberation in dedicated neural circuits. When above a certain threshold, such reverberation produces meaningful biophysical time bins in terms of specific temporal patterns which trigger, maintain and terminate a conscious brain state. Spatial information would be integrated into provisory topological maps at non-conscious levels through adaptive resonant matching, but not form part of the temporal access code as such. The latter, de-correlated from the spatial code, would operate without any need for firing synchrony on the sole basis of temporal coincidence probabilities in dedicated resonant circuits through the progressively non-arbitrary selection of specific temporal activity patterns in the continuously developing brain.

## 1. What is phenomenal consciousness?

The many different definitions of phenomenal consciousness proposed in the literature (e.g. Kihlstrom, 1987; Natsoulas; 1983; Dennett; 1991; Posner, 1994; Block, 1995; Revonsuo, 2000; Zeman, 2001; Dietrich, 2003) disappointingly reveal that a truly operational definition of the phenomenon as such, indispensable to its scientific investigation, has not been found, yet. The major problem here is that to define phenomenal consciousness, we refer to introspective considerations, as pointed out almost two centuries ago by William James (1890). In the first book (part 4, section 6) of the *Treatise of Human Nature* (1740), the Scottish Philosopher David Hume compared phenomenal consciousness to a theatre, a scene of complex events where various different sensations and perceptions make their successive appearance in the course of time:

> "The mind is a kind of theatre, where several perceptions successively make their appearance; pass, repass, glide away, and mingle in an infinite variety of postures and sensations. There is properly neither simplicity in it at one time, nor identity in different, whatever natural propension we may have to imagine that simplicity and identity. The comparison of the theatre must not mislead us. They are the successive perceptions only, that constitute the mind; nor have we the most distant notion of the places where these scenes are represented, or of the materials of which it is composed."

Hume's phenomenal description of successive perceptions appearing as sequences in time is embedded in some contemporary views of consciousness. Less than ten years ago, the neurobiologist Ramachandran discussed the concept of 'self' in relation with the concept of 'consciousness', and emphasized that phenomenal consciousness encompasses hardly more than sequences of many distinct perceptions and sensations (Ramachandran, 1998). This difficulty we seem to have in science to actually get a hold on what we call phenomenal consciousness compromises our attempts to work out a scientifically operational definition. A number of authors suggested that such attempts may be doomed in advance, or that we do not dispose of enough experimental data, yet (e.g. Searle; 1998, Crick & Koch, 2000; Humphrey, 2000). Dehaene and colleagues (Dehaene, Changeux, Naccache, Sackur, & Sergent, 2006) proposed a supposedly operational taxonomy for the scientific investigation of consciousness based on a distinction between subliminal, preconscious and conscious processing. While there is nothing new about this taxonomy as such, given that Kihlstrom (1987) already proposed exactly the same three-level model twenty years ago, Dehaene *et al's* distinction between vigilant states and what they call 'conscious report' highlights the somewhat sobering fact that the only means scientists have of knowing whether their human subjects are phenomenally conscious instead of merely being in a vigilant or wakeful state is when some event is reliably reported. This consideration may point towards a potentially operational definition of consciousness in terms of 'access of information to conscious report', already embedded in Block's more general concept of access consciousness (e.g. 1995), but it still poses a major problem. The immediate data of phenomenal consciousness may consist of multiple, rapidly succeeding events that are often not coherent enough (see Hume's argument given above) to be reported as accurately as the carefully controlled events in behavioural studies. More importantly, the contents of our consciousness are often totally disconnected from external events or stimuli otherwise there would be no such thing as imagination or creative thinking.

## 2. Where is phenomenal consciousness produced in the brain?

In the fourteenth century, well before the dawn of the Age of Enlightenment, some physicians were trying to find the *locus* of the human soul in the body. With the advent of modern functional imaging techniques, the localization of consciousness in the brain has become the pet subject of a small industry in contemporary science. Rapid technological progress promoting the development of imaging and electrophysiological techniques has made it possible to correlate cognitive function with increasingly precisely located neural activities and interactions in specific brain areas. The observation that various aspects of conscious information processing are correlated with local or global activities in the brain is anything but surprising, but such correlations neither explain phenomenal consciousness, nor do they help us explain or understand how the brain achieves to render non-conscious representations in long-term memory conscious at a given moment in time. Interestingly, Dehaene *et al.* (2006) claim that considerable progress has been achieved by contrasting brain activation images leading to conscious perception or not, some indicating that conscious activity appears to correlate with occipital neural activity, others pointing toward a correlation with late parieto-frontal activity (see Rees *et al*., 2002, for a review). At the same, the authors conclude that the results of such studies appear inconsistent and that no coherent picture has emerged from them. While they agree that it is important to design paradigms in which conscious perception is not confounded with changes in behaviour, in other words the observable stimulus-response system, their own approach does not avoid this trap. In the experimental work of Dehaene's group, consciousness is conceived as the result of stimulus-driven processing based on what neuroscientists refer to as 'attentional selection'. But is attention necessary for the brain to generate a conscious state? When we dream, i.e. in the total absence of external stimulation, we are not wakeful and therefore not attentive to anything, but we are phenomenally conscious of many things. Sometimes we may even be able to access and report these phenomenal data several hours later by recounting our dreams over breakfast.

Approaching the still missing link between the conscious mind and the human brain may require some renewal in our scientific minds. Maybe some new form of theoretically guided analysis of functional characteristics of neural activity (e.g. Buszaki, 2007) triggering a radical shift in our ways of thinking about consciousness is just what is needed now to produce a proper theory of how consciousness arises from the brain. For the time being, unequivocal evidence supporting any theory that claims to link phenomenal consciousness to the ways in which the brain works is still missing, and we have to make do with only pieces of the puzzle that constitutes our current knowledge of functional aspects of information processing by the brain.

## 3. Functional aspects of conscious and non-conscious information processing

The greatest part of the information processed by the brain is not made available to phenomenal consciousness (e.g. Gray, 2002). Velmans (1991) even suggests that almost all mental processes are of a non-conscious or pre-conscious nature and that consciousness is nothing more than a by-product. An even stronger claim comes from Pockett (2004), who argues that consciousness may only be epiphenomenal, which is somewhat reminiscent of Lashley's (1956) now famous statement that "no activity of mind is ever conscious". Other views consider that any mental process may operate either consciously or non-consciously, depending on prior knowledge, experience, or practice (e.g. Schneider & Shiffrin; 1977, Shiffrin & Schneider, 1977; Frith & Dolan, 1996; Baars, 1997; Ramsey *et al,* 2004). Kihlstrom (1987) emphasized that consciousness must not be identified with any particular perceptual-cognitive function such as stimulus discrimination, perception, memory, or the higher mental processes involved in judgment or problem-solving. Rather, consciousness would be an experiential quality that may accompany any of these functions. Some of them involve what is called ‘rationality’, a process or property of the mind that is not necessarily experienced consciously and would, according to Churchland (2002), have a skill-based nature. In fact, conscious processing mainly seems to consist of arranging the elements of knowledge retrieved at a given moment in time into a temporal sequence of “input” and “output” transfers necessary to execute thoughts or actions. The periods of pure thought in such a process may, according to Crick & Koch (2000), not be directly accessible to consciousness, as is frequently the case in the perception of specific properties of so-called illusory figures (e.g. Dresp & Fischer, 2000).

From a strictly functional point of view, there are not more than two properties of consciousness that would be consistent with most of the experimental evidence available and that most authors would probably agree upon: its rather limited information processing capacity, and its expression in terms of a unique, continuously refreshed and updated stream of processing within a limited temporal window (e.g. Duncan, 1980; Mangan, 2003; LeDoux, 2002; Dietrich, 2003). In terms of brain processing, conscious activity relies mainly on serial processing, which allows for only a very limited amount of information to be dealt with in a given time span. Most people cannot consciously follow two ideas at the same time, or consciously execute two even simple, simultaneous tasks (e.g. Cherry, 1953; Baars, 1998). This "conscious seriality" (Seth & Baars, 2005; Edelman, 2003) undeniably constrains any possible theory of consciousness. Non-conscious activity, on the other hand, is largely based on massively parallel processing and can therefore handle a lot more information (e.g. Mesulam, 1990; Hochstein & Ahissar, 2002; Mangan, 2003; Dietrich, 2003). The function of serialization in terms of an ordered list of conscious events (e.g. Page & Norris, 1998; Seth *et al*, 2006), discussed already half a century ago by Lashley (1951), is linked to the hypothesis that an event or piece of information, once made conscious, would become selectively available to other processes involved in producing thought and speech. This function of making non-conscious information accessible to the mind is an important achievement of brain evolution; the limited capacity of conscious processing, on the other hand, represents a major functional constraint, as highlighted by numerous psychophysical data which include the observations by Triesch, Ballard, Hayhoe & Sullivan (2003) showing that observers see sudden changes in visual scenes only and "just in time" when they need the information to solve a specific problem.

The limited capacity of conscious processing entails that it must entirely rely on working memory, which can handle the 'magic' number of about 7 representations (e.g. Vogel, Woodman, & Luck, 2001). Such a limitation severely constrains the top-down processes that can effectively operate within the temporal window of a conscious experience. As proposed earlier by Mangan (2003), the pre-conscious processes at the fringe of consciousness may provide some kind of buffer, which both compensates for and regulates the limited conscious capacity. The processing capacity of the non-conscious, in contrast, may be estimated within a range of at least $10^7$ bits, knowing that the optical nerve transfers $10^8$ bits per second (Koch, 1997), which is infinitely more than working memory can deal with, i.e. the well-known 7±2 items demonstrated by Oberly (1928), Miller (1956), and more recently by Parkin (1999). Treisman (1998) argued that there are too few neurons to individually encode the combinatorial explosion of arbitrary conjunctions that we are capable of processing consciously, taking as example that of a "purple giraffe with wings". This implies that 'purple giraffe with wings' is coded holistically once its representation has been internally validated, i.e. accepted by the mind as a 'thing that makes some kind of sense' to an individual. Also, we believe that whenever the brain builds a complex representation, it is inevitably matched to some past, present, or future event by a specific and purely temporal brain mechanism. How such temporal matching could be achieved will be explained later herein. In the course of time, a given match will either come up frequently and be consolidated (imagine, for example, an artist becoming obsessed by purple giraffes with wings), or it will eventually fade out. The limitations of conscious processing are defined in terms of the representational content "authorized" to invade the conscious workspace at a given time. Most of it would be concerned with complex objects or object relations and transformations. This complex material is retrieved from non-conscious long-term memory, where each integrated representation is 'tagged' by a specific temporal activity pattern. Temporary retrieval of a given 'tag' produces what we will as of now refer to as a 'conscious state'.

The notion of a conscious state as a potentially operational concept for the study of consciousness was successfully defended by Tononi & Edelman (1998) and encompasses the earlier definition by von der Malsburg (1997) in terms of a continuous process with a limited duration. A conscious state is not to be confounded with a state of awareness or vigilance (see also Nielsen & Stentstrom, 2005; Dehaene *et a.*, 2006). Although conscious states may involve cognitive processes such as memory (e.g. Cowan, Elliott, Saults, Morey, Mattox, Hismjatullina, & Conway, 2005), attention (e.g. Raz & Buhle, 2006), conscious report (e.g. Crick & Koch, 2000), or volition (Grossberg, 1999; Dehaene *et al.*, 2006), such implications will not be discussed here. Instead, we will focus on experimental data and theoretical arguments that further the conscious state notion as a scientifically operational concept. On the basis of this concept, we will bring to the fore how the pieces of the puzzle may eventually fit together and reveal a clearer picture of how brain mechanisms may trigger temporarily available conscious representations; these sequences of unique, successive events in time or, expressed in Humes' or Ramachandran's terms, 'successive perceptions', within a dynamic range of relatively brief durations.

## 4. Temporal brain mechanisms and conscious states

The limits of conscious states in terms of processing capacity and duration suggest, and may even impose, a temporal mechanism as the most parsimonious explanation for their genesis. The Lisman-Idiart-Jensen model has been the first to make an attempt in this direction (Lisman & Idiart, 1995, Jensen *et al*, 1996, Jensen & Lisman, 1996, Lisman, 1998, Jensen & Lisman, 1998, Jensen, 2005). It postulates that a temporal pattern code only is required to trigger and maintain a conscious state. While the conscious state may exploit spatial or topological information linked to the representations made available, these spatial contents remain encoded or 'tagged' at non-conscious levels.

Taking into account some of the experimental data and theoretical arguments discussed above, the Lisman-Idiart-Jensen model is composed of a working memory with a maximum processing capacity of 7±2 items. Each such item is represented by the firing of a specific cell assembly (the so-called 'coding assembly') during one gamma period, the whole phenomenon occurring in a theta period composed of approximately 7 gamma cycles. Detailed model accounts, for the slope of the Sternberg curve (38 ms per item), for example, were developed on the basis of this approach (Jensen & Lisman, 1998; 2005). Similarly, Başar (1998) and Başar *et al* (2000) considered cognitive transfer activities to be based on oscillations at *alpha, gamma, theta, delta* and other temporal frequencies which would be 'combined like the letters of an alphabet' to deliver a temporal code reflected through EEG and event-related potentials (ERP), analyzed in terms of wavelets.

Whether conscious states can be approached topologically, i.e. whether they occupy a precise and functionally delimited area in the brain like a particular prefrontal region, for example, or whether they involve long-range interactions between areas of the brain has been subject to debate (e.g. Dehaene & Naccache, 2001; Dehaene *et al*, 2003, Feinstein *et al*, 2004; Dehaene *et al.*, 2006). The heuristic value and merit of Lisman, Idiart & Jensen's model relies on the postulate that a conscious state can be triggered by a critical temporal activity pattern <u>anywhere</u> in the brain at any given moment in time. This is a radically different way of thinking about the brain genesis of consciousness because the latter is conceived in terms of a firing pattern independent of the functional identity of the cells that fire. A code for conscious state access thus would consist of a truly unique temporal pattern of activity retrieved for an individual conscious event, and regenerated whenever required by the same set of cells without any need for synchronous activity. Models developed by Helekar (1999) and John (2001) have provided theoretical and empirical arguments in favour of temporal codes for conscious state access.

These temporal codes are closely linked to the duration of a conscious state, or so-called ‘psychological moment’ (Pöppel & Logothetis, 1986; von der Malsburg, 1999; Tononi & Edelman, 1998), with variations in the limited dynamic range of a few hundreds of milliseconds. This has been established on the basis of a considerable body of psychophysical and neurobiological data (e.g. Lehmann *et al*, 1987; Lestienne & Strehler, 1988; Thorpe & Imbert, 1989; Crick & Koch, 1990; Potter, 1993; Strik & Lehmann, 1993; Gray, 1995; Pascual-Marqui *et al*, 1995; Taylor, 1996; Koenig & Lehmann, 1996; Lehmann *et al*, 1998; von der Malsburg, 1999; Bressler & Kelso, 2001; Chun & Marois, 2002). The work of Libet (1993; 2003; 2004), for example, has shown that a time minimum of about 500 ms is required for a near-threshold stimulus to produce a conscious perceptual experience. From an evolutionary viewpoint, the upper limit of a conscious state would correspond to some temporal duration beyond which a newly triggered conscious state, making representations potentially relevant to survival available, would be likely to come too late. In order to analyze neural patterns in terms of the temporal codes they deliver, the duration of a conscious state is to be divided into critical time windows, or 'bins', the length of which would be limited by the accuracy of neuronal timing, or the lower limit of biophysics. Such a time window, or ‘bin’, has been expressed by the parameter $\Delta t$ which would, in principle, represent the sum of standard deviations for the time delay of synaptic transmission including the duration of the refractory period. An average estimate of 6 ms for this parameter appears reasonable in light of the data available (Bair, 1999). Helekar (1999) based his calculations of a temporal code on an average duration of 3 ms for $\Delta t$, operating under the hypothesis of an average estimate of only 30 ms for a state duration, expressed in terms of the parameter $t$. An average value of 6 ms for $\Delta t$ would be consistent with ‘bin’ durations proposed by Shastri & Ajjanagadde (1993), Moore & King (1999), or Rieke *et al* (1997). Yoshioka & Shiino (1998) suggested 10 ms and Singer (2000) times no longer than 10 ms. Interspike intervals and integration times of cortical neurons are within a similar dynamic range (Eggermont, 1998).

Under the simple assumption that within each ‘bin’ there is either a signal or no signal, derived from McCullough & Pitts’ (1943) germinal work on information transmission in neural networks, the information content of each bin is 1 bit. On the basis of an average duration of 300 ms for a given conscious state, which seems more realistic than the 30 ms state duration suggested by Helekar, a 6 ms duration for a critical time window or ‘bin’ within that state, and with a deterministic signal being generated during each ‘bin’, the information content of such a conscious state would be 300/6 = 50 bits. A similar computation of the maximum quantity of information conveyed by a duration *t* with a number of temporal windows identified by a given $\Delta t$ was proposed by MacKay & McCulloch (1952). Considering there are equal probabilities for activity (signal) and non-activity (no signal) within each 'bin', a conscious state of a duration of 300 ms would then generate 61 bits of content (for $\Delta t$ = 6 ms). This theoretical approach is detailed in Rieke *et al* (1997), who also point out that actual neuronal systems approach the assumed theoretical limit of information transmission. The figures given above may be compared with estimates of the number of visual prototypes held in memory, given by Tsotsos (1990), which correspond to information contents of 17 to 23 bits. Similar time-based coding schemes were suggested later by Thorpe *et al* (2001) and VanRullen *et al* (2005). The biophysical code for conscious state access is defined in terms of critical temporal activity patterns that trigger and maintain conscious states. This code has the considerable advantage of operating independently from the functional identity of the neurons delivering it or the spatial 'tags' contained in the subjectively experienced events. The dynamic analysis of correlated oscillations in cortical areas at various frequencies (e.g. Bassett *et al*, 2006) and the study of functional interactions between gamma and theta oscillations in various structures of the brain (e.g. Axmacher *et al*, 2006) represent promising approaches here.

None of the existing neural models of cognition proposes a clear division between the functioning mode of processors or circuits generating conscious brain data from that of processors generating non-conscious events. In the light of the evidence considered above, we have serious doubts that any plausible functional account for human consciousness can be expected from a spatial model. If all the complex spatial brain data we may experience when in a conscious state were, indeed, part of the code for conscious states, then how would they be organized and reliably deciphered? A unifying, purely temporal code provides a ready and perfectly plausible answer here. The mechanisms for the generation of such a temporal code for conscious state access would be based on the functional characteristics of reverberating brain activity. Reverberation (Abeles *et al*, 1993; Edelman, 1993; Crick, 1994; Grossberg, 1999; Constantinidis *et al*, 2002; Lau & Bi, 2005; Dehaene *et al.*, 2006) would, among other things, explain how the immense variety of afferent input triggering a multitude of brain signals eventually leads to the critical temporal patterns for conscious state access. Reverberant neural activity has been studied in thalamo-cortical (Llinás *et al*, 1998; Llinás & Ribary, 2001; VanRullen & Koch, 2003) and in cortico-cortical pathways (Steriade, 1997; Pollen, 1999; Lamme, 2004; 2006). Reverberation is a temporal process that generates feed-back loops in the brain, described by various authors in terms of 're-entrant circuits'

(Edelman, 1989; 1993, Tononi *et al*, 1992; 1998, Tononi & Edelman, 1998; 2000, Edelman & Tononi, 2000; Fuster, 2000; Prinz, 2000; Di Lollo *et al*, 2000; Klimesch *et al*, 2001; Edelman, 2003; Robertson, 2003; Crick & Koch, 2003; 2005), and successfully implemented in fractal neural network models of the brain (e.g. Bieberich, 2002). Reverberation readily explains how the brain copes with situations where two or more representations with similar or identical probabilistic weights are trying to invade the conscious workspace at one and the same moment in time. Without reverberation, the conscious execution of focussed action would be difficult, if not impossible (e.g. Lamme, 2006). Two successive conscious states need to be separated by at least the time the brain needs to inactivate the current temporal access code and to generate the new one. If not separated from each other in time, the different sustained neural activities producing the critical temporal patterns would inevitably interfere with each other, like representations in short-term memory are annihilated by new input if they are not maintained through rehearsal (Potter, 1993).

## 5. Arguments for a temporal access code to consciousness

The idea of a purely temporal access code for consciousness as the most parsimonious link between brain and mind (Dresp-Langley & Durup, 2009; 2012) is compelling in the light of several theoretical arguments. If spatial coding took place within consciousness, the brain would have to integrate so many signals from multi-channel cross-talk that a reliable and unifying coding scheme seems almost inconceivable. We suggest that, for generating access to capacity-limited spatial and temporal representations within consciousness, the temporal code is de-correlated from the spatial code. De-correlation is an important notion in neural network theory and systems theory in general. It describes a mechanism that reduces crosstalk between multi-channel signals in a system (like the brain) but preserves other critical signal properties. As a reminder, we should like to emphasize that the term ‘code’ originally stems from information theory and may stand for both 1) an entire system of information transmission or communication (like the brain) where symbols are assigned definite meanings and 2) a set of symbols for the content of a given message (like a temporal activity pattern) within that system. One of the major arguments in favour of a purely temporal access code for conscious brain states would be its undeniable adaptive advantage.

### ***5. 1*** *Adaptive advantage and epigenetic plausibility*

Many authors have insisted on the important adaptive function of consciousness (Gray, 1971, 1995; Crick & Koch, 1995, Koch & Crick, 2000, de Charms & Zador, 2000). In line with their considerations and arguments, Helekar (1999) proposed a genetically programmed code for consciousness arising from non-arbitrary linkages between temporal firing patterns and subjective experience in a similar way as the genetic code arises from non-arbitrary linkage of anticodons and their cognate codons. Helekar was the first to fully realize that encoding conscious states by a single, unifying parameter would represent a considerable adaptive advantage compared with spatio-temporal codes, which would be far too complex and definitely more costly in terms of allocation of neural resources. Whether evolution has produced genes that drive conscious state generation, or whether this capacity would rather be a consequence of epigenetic development is subject to debate.

If there is such a thing as a genetic code for conscious states, then why did evolution not push the processing capacity of conscious states? Baars (1993) and Newman and Baars (1993) argued that nature would have calculated the trade-off between the benefits and the costs of increasing the processing capacity of consciousness, with the result that such an increase would have been too costly a process to justify its, assumedly minor, advantage. Such reasoning was countered by Pockett (2004), who argued that conscious capacity compared with the immense resources of non-conscious processes would represent only about 1/100.000 of brain capacity available, and that an increase in conscious capacity through evolution would not have been costly at all. We believe that, under the hypothesis of a purely temporal mechanism for conscious state access within a limited dynamic range of durations, not increasing the processing capacity of conscious brain states represents a sound strategy of evolution. For the sake of economic resource allocation, evolution has made a smart choice by only pushing non-conscious resources, i.e. the capacity of the brain to generate and integrate increasingly complex non-conscious representations. Making these fully integrated representations then available to consciousness remains the sole task of the temporal access code, with no need to increase conscious processing capacity any further. Also, during ontogenetic brain development representations remain largely non-conscious for a long time before some of them eventually become the subjectively and holistically experienced data of a human being's phenomenal consciousness, at the age of two or three. Sensory, somatosensory, and proprioceptive signals may instantly be perceived as the immediate data of a conscious state, eliciting what psychophysicists call sensations, but these are continuously integrated into representations by non-conscious mechanisms. This, among other things, explains the striking similarities between descriptions of objects resulting from direct perception and from pure imagination (Kosslyn, 1994; 1999; Kosslyn *et al*, 2001). Because of their considerable adaptive advantage as well as the adaptive properties of the brain mechanism underlying their genesis, which we will discuss later, we suggest that the origin of the temporal activity patterns for conscious state access is epigenetic.

### ***5.2*** *Functional plasticity and Ramachandran's spatial 're-mapping hypothesis'*

The integration of signals originating from the different sensory modalities into topologically coded representations has to be sufficiently adaptable and it has to display a certain functional plasticity to enable the continuous updating of these representations as a function of changes in contents. Such changes are imposed on our brains day by day by new situations and experiences. Yet, to be made available to consciousness, there has to be some permanently reliable, unifying "tag" for these continuously updated, re-integrated representations to ensure a stable access over time irrespective of changes in contents. Grossberg (1999) referred to this problem as the *"plasticity-versus-stability dilemma*" and proposed resonant brain learning mechanisms as a potential solution. While these latter satisfactorily resolve the dilemma at the level of non-conscious information processing, they fail to explain how non-conscious representations would become available to consciousness. By the vague claim that not all resonant brain mechanisms generate conscious representations, but that all conscious representations would be based on resonant mechanisms, without specifying the functional characteristics that would generate the passage from one to the other, Grossberg re-introduces the *plasticity-versus-stability dilemma* at the level of the functional transition between non-conscious and conscious states. This particular point will be discussed further in our chapter on *'Adaptive resonance and Grossberg's dilemma'*.

Some compelling observations which reveal the extraordinary plasticity of spatial and topological coding in the brain have been reviewed by Ramachandran (1998). These include the author's own experimental work on so-called 'phantom limbs' (e.g. Ramachandran, Rogers-Ramachandran, & Cobb, 1995), a phenomenon first described in 1872 and repeatedly observed in hundreds of case studies since. After arm amputations, patients often experience sensations of pain in the limb that is no longer there, and experimental data show that a third of such patients systematically refer stimulations of the face to the phantom limb, with a topographically organized map for the individual fingers of a hand. On the basis of similar evidence for massive changes in somatotopic maps after digit amputation, and other experimental data showing that several years after dorsal rhizotomy in adult monkeys, the region corresponding to the hand in the cortical somatotopic map of the primate's brain is activated by stimuli delivered to the face (Merzenich *et al*, 1984), Ramachandran and his colleagues proposed their 'remapping hypothesis' (e.g. Ramachandran, Rogers-Ramachandran & Stewart, 1992). The latter clarifies how spatial and topological representations are referred to other *loci* in the brain through massive cortical re-organization. The findings reported by Ramachandran and others deliver compelling elements of proof that, despite dramatic changes in non-conscious topology, representations remain available to conscious state access and may still experienced in terms of sensations such as pain, cold, digging or rubbing. This phantom-like persistence of conscious representation in time but not in space is possibly one of the strongest arguments for conscious state access through a temporal code independent of, or de-correlated from, the spatial code.

### ***5.3*** *The temporal 'coherence index' and coincidence detection*

In his 'neurophysics of consciousness', John (2001, 2002) suggests that a conscious state may be identified with a brain state where information is represented by levels of coherence among multiple brain regions, revealed through coherent temporal firing patterns that deviate significantly from random fluctuations. This assumption is consistent with the idea of a reliable temporal code for conscious state access despite spatial remapping through cortical re-organization. Empirical support for John's theory comes from evidence for a tight link between electroencephalographic activity in the gamma range defined by temporal firing rates between 40 and 80 Hz (i.e. the so-called '40-Hz' or 'phase-locked' gamma oscillations) and conscious states (e.g. Engel *et al*, 1992). This 'coherence index', with its characteristic phase-locking at 40 Hz, was recently found to change with increasing sedation in anaesthesia, independent of the type of anaesthetic used (Stockmanns *et al*, 2000), with decreasing temporal frequencies when doses of a given anaesthetic were increased. Moreover, the characteristic phase-locking at 40 Hz displays coherence not only across brain regions during focussed arousal, but also during REM sleep, when the subject is dreaming (Llinás & Ribary, 1993). Coherence disappears during dreamless, deep slow-wave sleep, which is consistent with the findings reported on deeply anesthetized patients referred to above. The fact that the temporal coherence index of a conscious state is produced during focussed arousal as well as in dream states is completely consistent with the view that dreams, imaginations, or daydreams represent genuine conscious states in the absence of wakefulness and external trigger stimuli.

The phase-locking at the critical temporal frequency would be achieved through intra-cortical reverberation, enabled by a digital event within a hybrid system, according to John's terminology (John, 2001, 2002). This hybrid system, the brain, establishes non-random departures from different *loci* or topological maps. These latter may undergo functional re-organization, yet, the temporal code for conscious state access remains coherent. This would lead to cortico-thalamic feedback loops, or resonance loops which generate the temporal access codes for conscious states through probability-based detection of memory events coinciding in time. The mechanisms which would explain how memory events are read out and compared in the brain were discussed by Grossberg in his Adaptive Resonance Theory (1975; 1999).

### ***5.4*** *Adaptive resonance in the continuously learning brain*

Originally, Adaptive Resonance Theory (ART) was conceived as a theory of brain learning to explain how the brain generates and updates representations of continuously changing physical environments (Grossberg, 1975). More recently, ART was extended to account for related phenomena such as attention and intention or volition. According to Grossberg (1999), the link between these three could be described by the fact that intentions would lead to focus attention on potentially relevant internal or external events. These *foci* of attention would lead to new representations when the system (the brain) is able to validate and integrate them into resonant states, which would include, according to Grossberg, the conscious states of the brain. According to the theory, all conscious states would be resonant states, triggered either by external or internal events and mediated by either attention or volition. Thisnas such, however, does not explain how non-conscious representations would become available to ongoing consciousness. In our analysis, this is a direct consequence of the fact that the theory fails to separate spatial from temporal coding and thereby fails to resolve Grossberg's *stability-versus-plasticity* dilemma at the level of the transition from non-conscious representation to conscious state access (see chapter *6.2* above). Adaptive resonance theory nonetheless plausibly explains how the brain ensures the continuous updating of non-conscious representations through a mechanism termed top-down matching, which produces the so-called resonant brain states.

A resonant brain state would be achieved through the repeated matching of external or internal events in short-term or working memory to internal events activating top-down representations. According to the theory, the brain is continuously confronted with ongoing internal or external representations (*bottom-up*) and therefore has to continuously generate probabilistic hypotheses to determine what all these transitory events are most likely to be and whether they are relevant. This involves matching the ongoing representations to representations stored in long-term memory (*top-down*). Coincidence of bottom-up representations and top-down representations (*top-down-matches*) would produce so-called matching signals, or coincidence signals which, when repeatedly generated, lead to resonant states in the brain. The representations generated through top-down matching of brain signals would be, according to Grossberg, coded topologically in the 'What' and 'Where' processing streams of the brain (see Grossberg, 1999 for an extensive review of relevant physiological data), and what he calls "the resonant code" is therefore tightly linked to functional topological organization. The question how non-consciously encoded topological information would be made available to consciousness is left unanswered. We propose that the resonance states corresponding to conscious states arise from temporal integration only, given that only non-conscious states would dispose of enough capacity to integrate signals across both time and space. How reverberation of temporal signals, probabilistic signal coincidence detection and adaptive resonance in dedicated neural circuits would produce what we call the temporal access code for conscious states is explained in the following chapter, which introduces our own model.

## 6. The 'time-bin' code

While there is no empirically based description of resonators receiving, amplifying and transmitting time-patterned messages in the brain, it is nevertheless certain that a large number of physical and biophysical phenomena can be plausibly and parsimoniously explained on the basis of resonance principles or mechanisms. We believe it makes good sense that evolution (see *6.1* above) would have produced brains capable of generating conscious states on the basis of resonance mechanisms. How this may work, is shown in our model here.

### ***6.1*** *Underlying assumptions*

It is likely that biological resonators, in contrast to "ordinary" resonance devices designed by humans, would almost certainly have highly sophisticated operating principles, given that hundreds of functionally different kinds of cells exist in the brain. On the other hand, there is no reason why resonators in the brain would have to function with a high level of precision, provided they operate according to some redundancy principle and the whole ensemble of cells producing a conscious resonance state behaves in a statistically predictable way. Our model conception of temporal signal sequences forming a specific biophysical ‘time-bin’ pattern that activates, maintains, and inactivates a conscious state is certainly and inevitably a simplification of reality. Such a simplification does, however, not affect the internal validity of the model arguments presented here. The principal aim of our model is to explain how a ‘time bin resonance system’ would generate conscious brain states on the basis of a relatively limited amount of neural resources.

Given the known temporal properties of conscious information processing, we suppose that conscious states may generate messages corresponding to variable contents in terms of bit sequences corresponding to variable durations. Any of these conscious states would be identified by a unique sequence of 1s and 0s. Thus, in the same way as bar codes provide the key to an almost infinite variety of things, such temporal sequences provide the key to consciously experienced brain events. A given temporal code would be generated spontaneously at a given moment in early brain development and would then eventually be reproduced and consolidated during brain learning (see Figure 2). Consolidation would be a result of repeated reverberation in cortical memory circuits, leading to resonance states which correspond to more or less specific conscious states. Once a resonance circuit has been consolidated for a given temporal sequence, a resonant state is automatically activated whenever there is a statistically significant temporal coincidence between representations at a given moment in time. As long as the threshold of statistically significant coincidence is not attained, representations in the resonance circuit remain non-conscious or pre-conscious.

Counting from a first signal, or spike, in biophysical time, the resulting temporal sequence of 1s and 0s may be described as a succession of intervals ($q$) between 1's. Let us imagine a network of brain cells, or a resonator, with a functional architecture or connectivity described by the shapes of closed polygons (see Figure 1 for an illustration), with a variable number $q$ of apices and the same number of edges. Each apex of such a polygon would correspond to a neuron which can receive or emit input or output signals from and to processors anywhere in the brain as well as along the specific tracks of the resonant circuit that was primed during brain development for a specific temporal pattern tagging a conscious state. Here, we refer to the apices of our network model in terms of dedicated *principal resonant neurons*. Each edge of a polygon would represent a *delay path* which transmits signals from a given apex to the next, with a characteristic delay corresponding to some multiple of the elementary ‘bin’ unit (Δt, as defined earlier by others in other models discussed earlier here). The distribution of these delays should fit the proportion of 1's and 0's in typical ‘time-bin’ messages: if, for example, 1's are as likely to occur in a code as 0's, then the proportions of various delays Δt, 2Δt, ..., *n*Δt would be predictable. The delay paths as such would correspond to local neural architectures in the brain. Whatever the effective operational structure of such a resonance circuit, its specific temporal characteristics would be experience-dependent and consolidated during development.

A brain or system operating on the basis of purely temporal resonance principles would work as follows. All principal resonant neurons would have been primed during brain development to preferentially process statistically significant signals. Thus activated, principal resonant neurons would send signals along all delay paths originating from them, and all those receiving a signal coinciding with the next input signal would remain activated. The connections between principal resonant neurons of the circuit would thereby be potentiated, as in the classical Hebbian model. Simultaneously, signals travelling from initially activated neurons to connected cells with too long delay paths would be cancelled. Thus, once a given polygon of a resonant network is potentiated along all of its edges, it would reverberate temporally coinciding signals while amplifying more and more the potentiation of the resonant connections. Now, let us consider the example of a simple sensorimotor task, which can be performed either consciously or non-consciously. Obviously, the message sent by the sensory system has to be decoded by the motor system. This would happen via non-conscious signal exchanges between functionally specialized neuronal assemblies. A conscious state, where the content of the representations activated by these signal exchanges between functionally specialized systems in the brain becomes subjectively experienced data of consciousness, is only triggered if the temporal coincidence between signals reverberating within resonant circuitry generates levels of potentiation beyond a given statistical threshold. How neuronal circuits would be able to learn statistical temporal information embedded in distributed patterns of activity was recently discussed by Gutig & Sompolinski (2006).

A network model of dedicated temporal resonance circuitry with a polygone shaped architecture may also generate quantitative predictions, as could be demonstrated by numerical simulations. If we consider, for example, a model structure with only 10 000 principal resonant neurons, each connected to only 5 others, numerical simulations would show that we then would have a resonant circuit that is able to reverberate signals, contents, or messages coding for biophysical time spans up to 50 bin. Whether a resonant circuitry reverberating bin codes for conscious state access would be localized or distributed all over the brain becomes irrelevant in regard to the probabilistic temporal coincidence hypothesis. It seems plausible, and likely to us, that inter-connected resonant circuits would develop all over the cortex during lifespan brain learning.

### ***6. 2*** *Developmental selection of non-arbitrary temporal activity patterns*

Like time-bin resonance itself, the selection of the critical temporal firing patterns that constitute the access code for conscious states use purely statistical criteria leading to fewer and fewer consolidated patterns for increasingly complex and integrated signal coincidences as our brain learns and develops. When we are born, all brain activity is more or less random. During brain development, temporal activity patterns elicited by events in biophysical time (*t*) ranging from 30 to approximately 500 ms (as explained above) will be linked to particular conscious experiences in a decreasingly arbitrary manner as frequently occurring, highly likely ('relevant') codes are progressively consolidated through a process called 'developmental selection'. This is illustrated in Figure 2, which is our adaptation of Figure 6 from Helekar's (1999) original paper. The model approach we propose here thereby resolves a critical problem in Helekar's model by explaining how non-arbitrary linkage of codes and contents is put into place progressively by developmental processes within dedicated resonant circuits of the brain. These developmental processes would operate on the basis of selective matching with a statistic temporal coincidence criterion, as explained above. In fact, once a given temporal code has been arbitrarily linked to a conscious state, it would remain potentially available as a 'brain hypothesis', which may then be repeatedly confirmed and consolidated or not. Once consolidated, the linkage of a code to content is non-arbitrary, or deterministic.

### ***6.3*** *From temporal activity patterns to dynamic resonant coding*

In his model, Helekar proposes a one-to-one non-arbitrary linkage between elements of subjective experience and specific temporal activities of neuronal assemblies. Non-arbitrary linkage would be, according to Helekar, innate and genetically driven. The mechanisms that execute these non-arbitrary operations are, as pointed out by Helekar himself, unknown. Once again, we find ourselves confronted with theoretical reasoning in terms of some kind of obscure superstructure. To overcome this problem with Helekar's model, we propose a selection mechanism that would operate on the basis of probabilistic learning in the memory circuits of the brain during lifespan development, progressively leading to deterministic and non-arbitrary temporal activity patterns for conscious state access. Helekar's "elementary experience-coding temporal activity patterns" were conceived in terms of a subset of neural firing patterns belonging to the set of all possible temporal patterns that can be generated by the brain. The original hypothesis states that only those patterns that are members of the subset would give rise to elementary subjective experiences, or conscious states, upon their repeated occurrence; repeated occurrence of so-called non-coding patterns would not give rise to conscious states. The problem with this approach is that the contents we may subjectively, or consciously, experience are also represented non-consciously in the brain. Helekar suggested that it would be the subjective nature of phenomenal consciousness *per se* that is genetically determined, which brings us right back to the "old" question, pointed out by Hume centuries ago and cited above: what *is* phenomenal consciousness? To avoid this old trap, our model proposes that conscious state access is based on 'time-bin' patterns corresponding to a temporally deterministic resonance state in dedicated circuits. How such states are consolidated outside consciousness through the repeated matching of current representations to representations in long-term memory is explained in Grossberg's theory (1999), discussed earlier here.

### ***6.4*** *From dynamic resonant coding to biophysical eigenstates*

What distinguishes a conscious state from a non-conscious state in our model would solely depend on a probabilistic criterion. A brain mechanism achieving coincidence computation would lead to the activation of a given temporal resonance code at a given time on the basis of a statistical coincidence criterion, or *coincidence threshold*. While Helekar (1999) suggested that the subjective nature of the conscious experience *per se* would determine an innate and genetically pre-wired temporal code, we propose a brain mechanism that would produce such a code through resonant learning in terms of temporal matches independent of the subjective nature of phenomenal consciousness. A conscious state would arise from a temporarily retrieved resonance state, tagged by a specific temporal pattern and generated within reverberating neural circuits which are updated outside consciousness during lifespan brain development. In fact, what is called "experience" in common language is re-coded in the brain in terms of temporal signal sequences in purely biophysical time.

The statistical coincidence of specific temporal sequences would activate, maintain, and inactivate conscious states in the brain like a bar code activates, maintains, or inactivates the electronic locks of a safe. Given the almost infinite number of signal sequences that are possible in such a code, there is no reason why there should not be a unique temporal pattern for a unique conscious state. In terms of quantum physics analogy, the time-bin resonance model suggests that non-conscious states are described by temporal wavefunctions which do not have a well-defined period. While a non-conscious state may be a combination of many non-specific *eigenstates*, resonant activity beyond the probabilistic coincidence threshold produces the well-defined temporal activity pattern or wavefunction of a single specific *eigenstate,* the '*conscious eigenstate'*.

## 7. Conclusions

A large part of the recent theoretical and empirical work devoted to the question of consciousness has consisted of trying to correlate conscious mental representations with neural activity in specific regions of the brain. Questions about a mechanism for the brain genesis of conscious activity have been neglected and there still is an immense explanatory gap between subjective conscious experience and brain functions. The problem at the root of this explanatory gap has become quite clear: the immediate data of phenomenal consciousness are too complex and often totally disconnected from any external stimuli or events to provide the basis for a scientifically operational definition of consciousness. In this article, we present an analysis that focuses on functional constraints of so-called conscious states, in line with definitions proposed earlier by others (e.g. Tononi & Edelman, 1998). We then show how the limited processing capacity and the unique stream of information processing in time that universally describe conscious states lead the way to a scientifically operational definition, which encompasses what we refer to as 'the time ordering function' of conscious states, where complex past, present and future events are represented in one and the same moment of conscious time. On the basis of a selective and thorough review of relevant experimental and theoretical data on temporal characteristics of conscious states, we propose a theory of conscious state access based on an epigenetic, biophysical code. The latter would be composed of resonant temporal activity patterns in the brain, generating time bins of several milliseconds each, with a few hundreds of milliseconds for a given conscious state. The probabilistic coincidence of resonance signals in time only provides the unique brain activity patterns that trigger, maintain, and terminate a conscious state like a bar code activates, maintains, or inactivates the electronic locks of a safe. The specific temporal patterns for conscious state access would be consolidated during brain development.

This suggests that our conscious brains become connected with the physical world in the course of their development, which is to be conceived in terms of a lifelong process. In a way similar to that of sonar systems which connect to the outside by acquiring some form of knowledge of a physical environment, conscious states are encoded in our brains in terms of some critical temporal base frequency as through scanning or pulsing. Although a conscious state may be experienced in any form of psychological space or time, the associated biophysical periods in the brain 'scale' this experience through a completely self-sufficient code. This explains how the inner clocks of consciousness can operate independently from spatial, verbal or any other form of cognitive or emotional experience. The brain does not care about the "exciting", "creative", "active", "boring", or "passive" subjective nature of conscious experience, which may lead to variations in subjectively experienced time later recalled as "time was flying by" or "time was standing still" (see Figure 4), it contents itself with scaling the signals produced by such events in its own, biophysical time.

The temporal code model addresses the mind-body problem at its very root. Some time ago, Nagel (1974) insisted that, in order to understand the hypothesis that a mental event is a physical event, we require more than the understanding of the word 'is', and that what we need most would be some plausible idea of how a mental and a physical term might refer to one and the same thing. Here, we have proposed such an idea and taken the risk of simplifying the question of consciousness at a moment where neuroscientists are struggling with a mass of evidence for complex correlates between consciously reported perceptions and spatio-temporal firing activity in functionally specialized cortical areas. In the ever thickening forest of facts and conjectures, a few will eventually stand out and become landmarks on the path of the science of consciousness. The fascinating experiments by Ramachandran and colleagues (e.g. Ramachandran, 1998), revealing the independence of conscious sensation from constraints imposed by topological cortical organization, supports the idea that conscious state access is not generated by any spatio-temporal code. The recent theoretical work by Lazar, Pipa, & Triesch (2007) on interactions between spike timing dependent and intrinsic synaptic plasticity in recurrent neural networks for the dynamic genesis of specific sequences or series of temporal activity patterns points towards future developments that may further our understanding of the functional characteristics of temporal brain codes. As for now, a temporal access code de-correlated from topological functional organization seems the most parsimonious explanation of how consciousness arises from brain function on the basis of mechanistic principles that take into account what we believe to have learnt about the brain.

Figure 1: Genesis of resonance states in a dedicated circuit with five principal resonant neurons acting as 'coincidence detectors'

a)

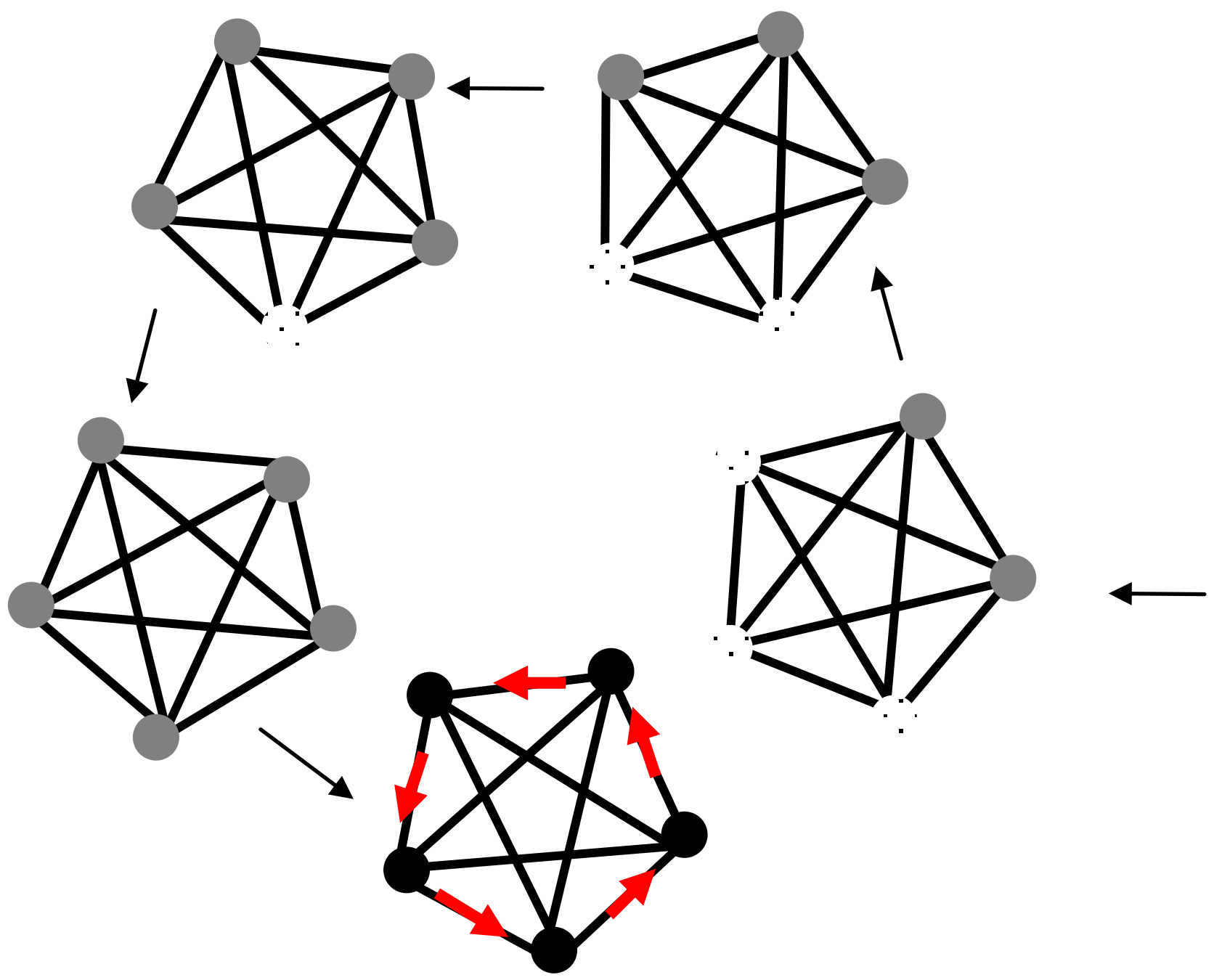

b)

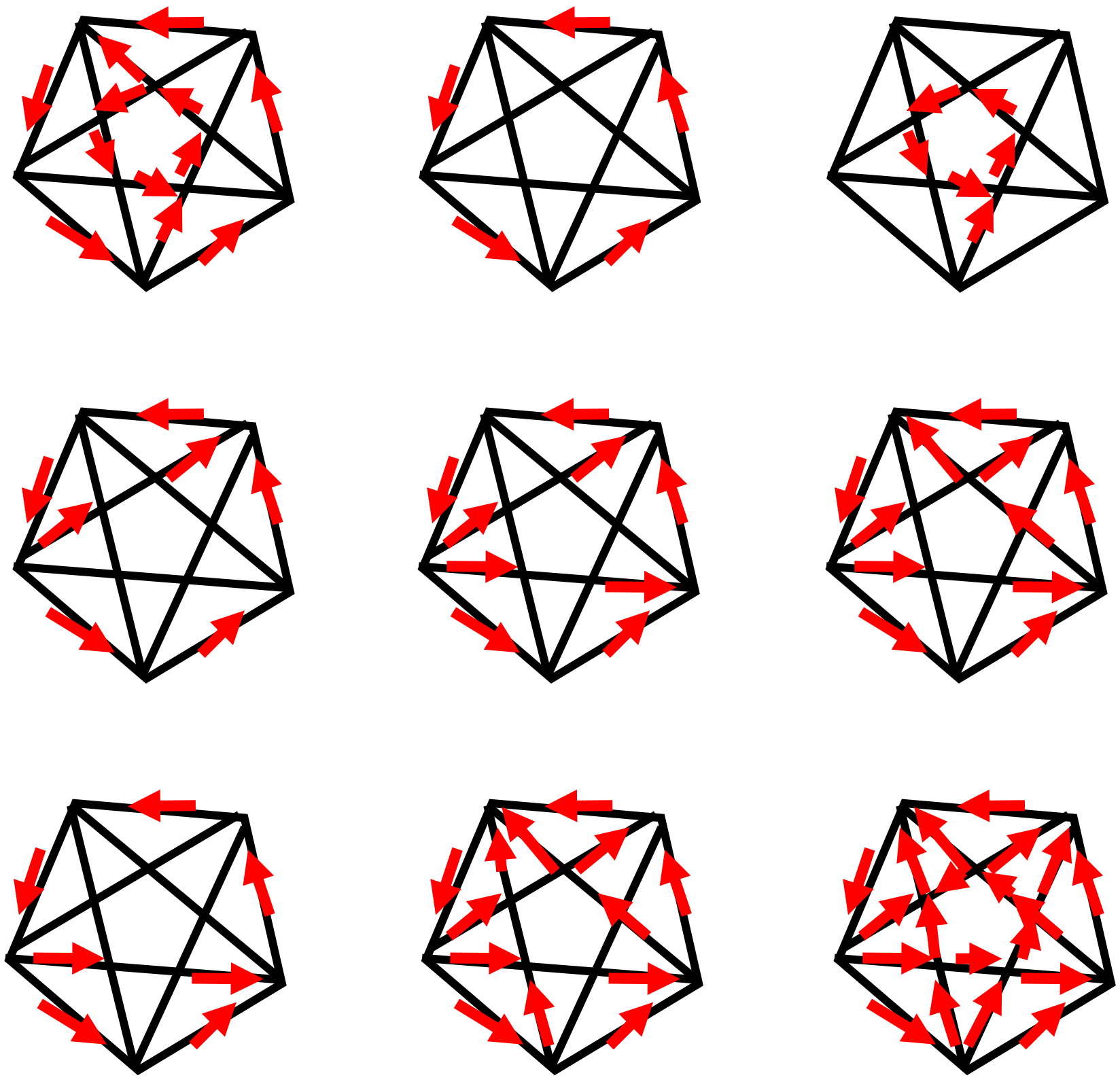

Figure 2: Developmental selection of temporal activity patterns coding for conscious state access

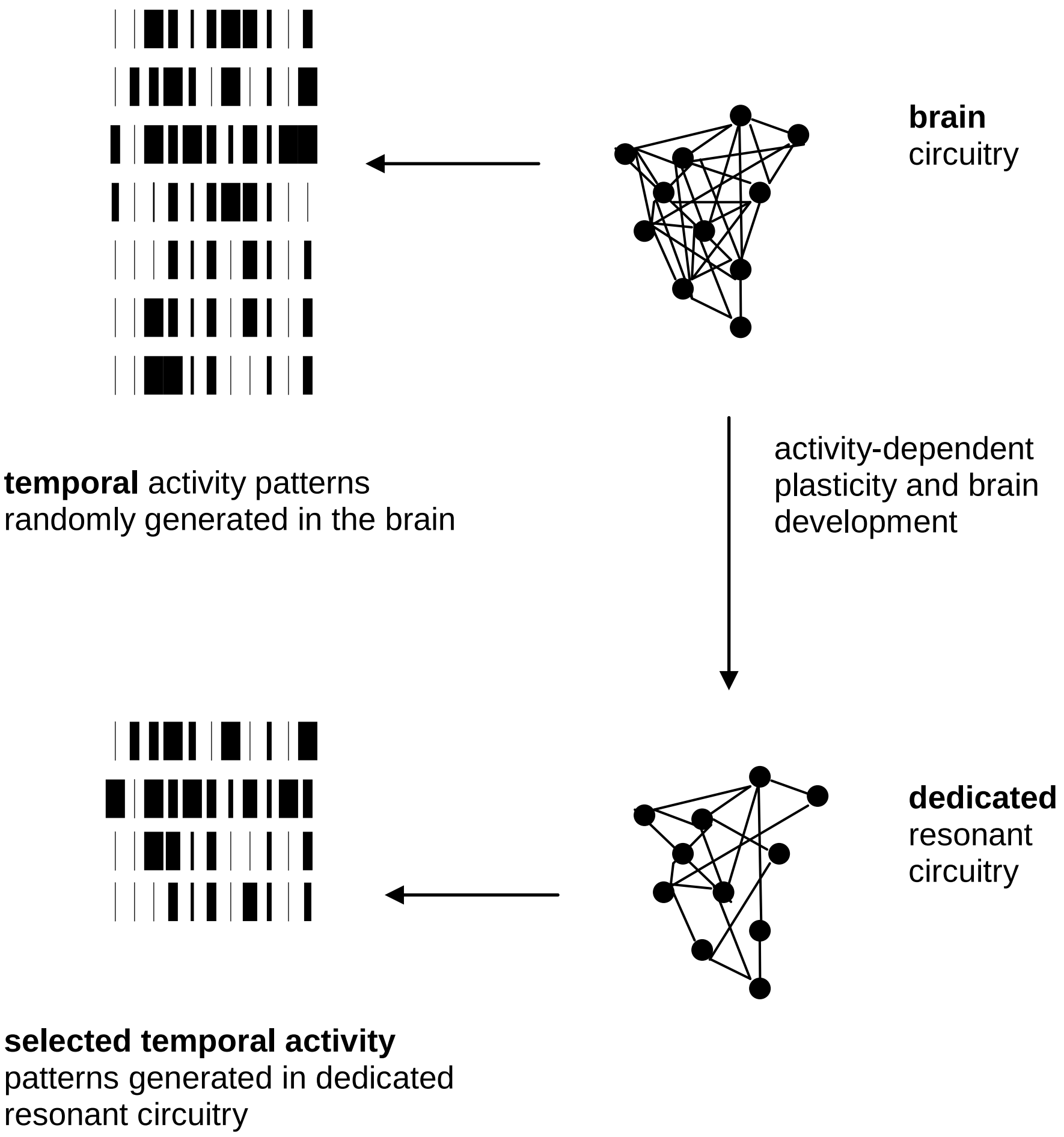

<u>Figure 3</u>: "Top-down matching" (after Grossberg, 1997, 1999) generates resonant brain activity for non-conscious memory representation at a given moment in time

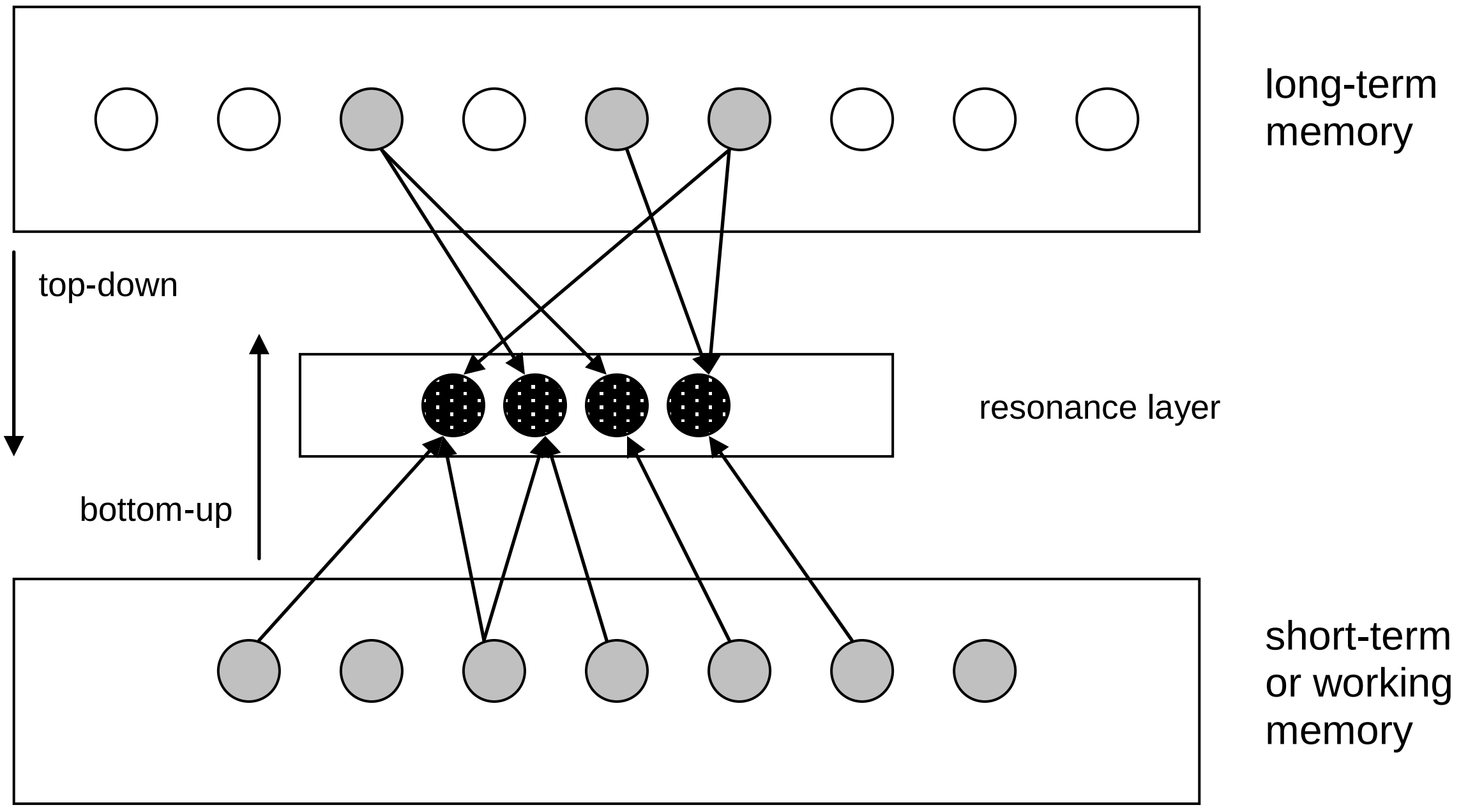

Figure 4: The conscious *eigenstate* (after Dresp-Langley and Durup, 2009) as a function of biophysical and subjectively recalled time

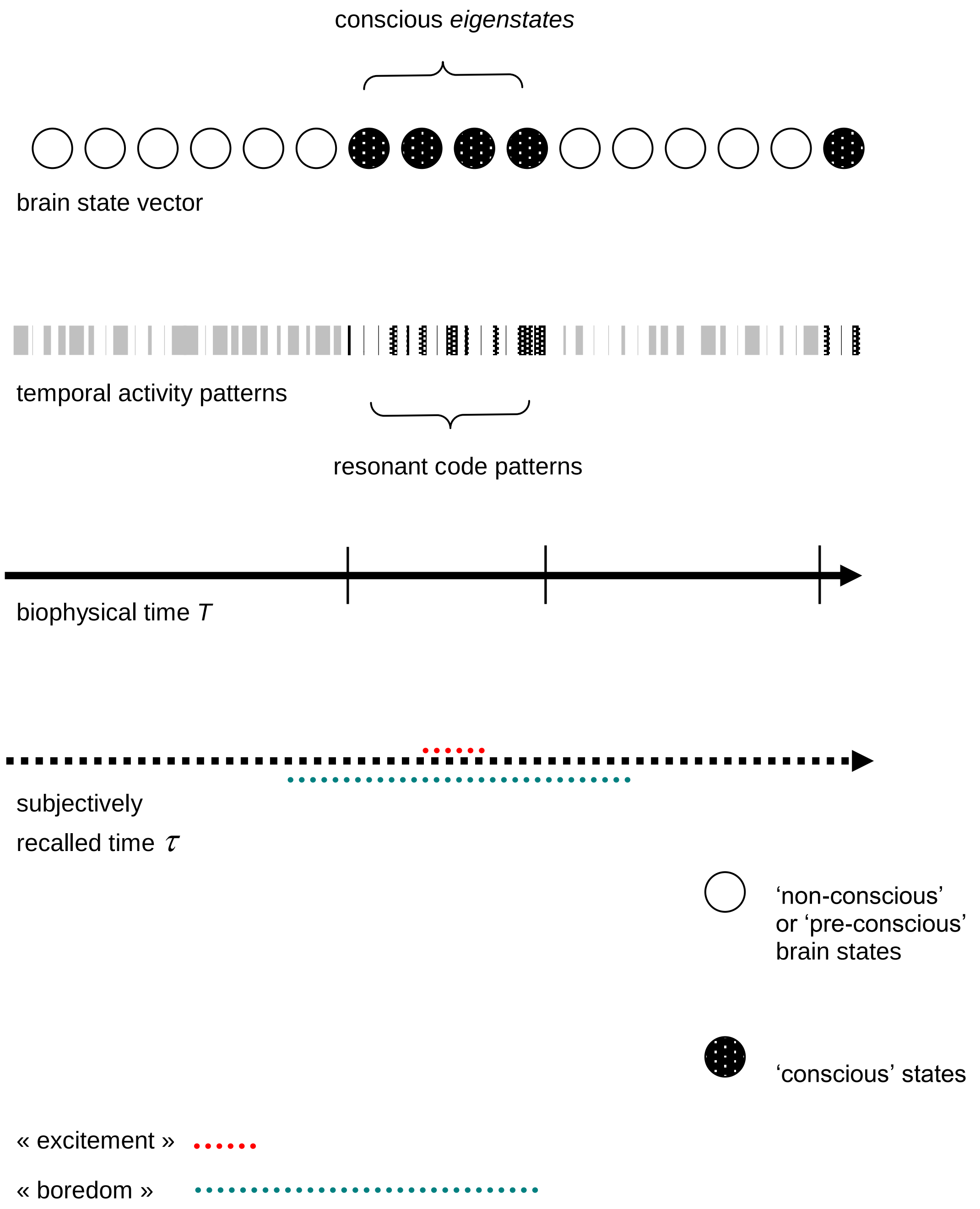

## Figure captions

<u>Figure 1</u>

Figure ***1a*** illustrates how a dedicated resonant circuit with five principal resonant neurons acting as coincidence detectors may be formed. Each apex of a given polygon would correspond to a *principal resonant neuron* which can receive or emit input or output signals from and to processors anywhere in the brain as well as along the specific tracks of the resonant circuit that has been primed in the course of brain development for a specific temporal pattern signalling for a conscious state. Unidirectional priming only is shown here, for illustration. Each edge of a polygon would represent a *delay path* which would transmit signals from a given apex to the next, with a characteristic delay that would correspond to some multiple of the elementary ‘bin’ unit ($\Delta t$, as explained in section 5.3). All principal resonant neurons would have been primed throughout lifespan brain development to preferentially process input which carries statistically ‘strong’ signals. Thus activated, principal resonant neurons would send signals along all delay paths originating from them, and all those receiving a signal coinciding with the next input signal would remain activated. The connections between principal resonant neurons of such a model would be thereby *potentiated,* like in the classic Hebbian model. Figure ***1b*** shows some of the many possible excitation patterns within a dedicated resonance circuit with only five principal neurons.

<u>Figure 2</u>

Figure 2 illustrates schematically how the critical temporal activity patterns for conscious state access would be progressively selected through activity dependent plasticity during lifespan brain development. At birth, a potentially infinite number of temporal activity patterns would be generated more or less randomly in the neural circuits of the brain. As brain learning progresses, repeated bottom-up-top-down matches (see Figure 3) of current brain events to learnt memory representations would generate resonant states in reverberating circuits which then progressively become dedicated resonant circuits. Whenever the firing patterns produced by these dedicated resonance circuits reach a statistical *temporal coincidence threshold,* the temporal firing pattern generated then would activate a conscious brain state. Thus, the temporal code of our ‘time bin resonance’ model would unlock the door to consciousness in a similar way as some bar code would unlock the door of an electronically protected safe.

Figure 3

Figure 3 illustrates schematically how Grossberg's Adaptive Resonance Theory (1999) accounts for the matching of bottom-up signals generated by current events to top-down signals generated by representations activated in long-term memory. This matching process is termed "top-down matching" and explains how non-conscious representations may be updated at any given moment in time via resonant circuitry in the brain.

Figure 4

Figure 4 illustrates how a conscious *eigenstate* of the brain may be conceived as part of a state vector as a function of biophysical time ($T$) and subjectively recalled time ($\tau$). In our model, the duration of a conscious *eigenstate* would correspond to a given number of biophysical 'time bins'. Biophysical time ($t$) is independent of the subjectively recalled duration of a given experience by a human individual, and would correspond to the duration of the critical temporal activity pattern produced by dedicated resonant circuits (Figure 1) to activate, maintain and inactivate a conscious *eigenstate*. Our 'time bin model' thus explains how the inner clocks of consciousness can operate independently from subjective experience, where variations from "interesting" to "dull" may produce variable, subjectively recalled durations of events.